\documentclass[%
 aip,
 jmp,%
 amsmath,amssymb,
 reprint,%
]{revtex4-1}

\usepackage{bibentry}
\usepackage{graphicx}% Include figure files
\usepackage{dcolumn}% Align table columns on decimal point
\usepackage{bm}% bold math
\begin{document}

\preprint{AIP/123-QED}

\title{Symmetric set of transport coefficients for collisional magnetized plasma}

\author{James D. Sadler}
\affiliation{%
Los Alamos National Laboratory, Los Alamos, New Mexico 87545, USA
}%
\author {Christopher A. Walsh}
\affiliation{%
Lawrence Livermore National Laboratory, Livermore, California 94550, USA
}%
\author {Hui Li}
\affiliation{%
Los Alamos National Laboratory, Los Alamos, New Mexico 87545, USA
}%

\date{\today}% It is always \today, today,

\maketitle

\textbf{
 In Braginskii extended magneto-hydrodynamics (ExMHD), applicable to collisional astrophysical and high energy density plasmas, the electric field and heat flow are described by the $\alpha$, $\beta$ and $\kappa$ transport coefficients. We show that magnetic transport relies primarily on $\beta_\parallel-\beta_\perp$ and $\alpha_\perp-\alpha_\parallel$, rather than $\alpha_\perp$ and $\beta_\perp$ themselves. However, commonly used coefficient fit functions  [Epperlein and Haines, Phys. Fluids \textbf{29}, 1029 (1986)] cannot accurately calculate these quantities. This means that many ExMHD simulations have significantly over-estimated the cross-gradient Nernst advection, resulting in artificial magnetic dissipation and discontinuities. We repeat the kinetic analysis to provide fits that rectify this problem. Use of these in the Gorgon ExMHD code resolves the known discrepancies with kinetic simulations in the literature. Recognizing the fundamental importance of $\alpha_\perp-\alpha_\parallel$ and $\beta_\parallel-\beta_\perp$, we re-cast the set of coefficients to find that each of them now shares the same underlying properties. This makes explicit the symmetry of the magnetic and thermal transport equations, as well as the symmetry of the coefficients themselves.}

Treatment of collisional magnetized plasma with the electron-ion two-fluid approach leads to a theory of magnetic transport \cite{braginskii1958transport} as a function of the fluid properties. This collisional extended magneto-hydrodynamic (ExMHD) theory is based on the assumption that, since electrons are much lighter than the ions, they will quickly form a sheath around the ion fluid. The electric field $\mathbf{E}$ of this sheath leads to transport of the magnetic field. In ideal MHD, $\mathbf{E}=0$ in the fluid rest frame. This implies that the magnetic field $\mathbf{B}$ is simply advected with the fluid flow, although advection along $\mathbf{B}$ has no effect. 

Other processes in the electron momentum equation, however, lead to greater complexity. Coulomb collisions give rise to Ohmic resistance. Electron temperature gradients produce  thermoelectric forces, since hotter electrons are less susceptible to collisions. Subsequently, it was recognized \cite{braginskii1958transport} that the resistive and thermoelectric processes should be described by tensors dependent on the direction of $\mathbf{B}$. Typically, ExMHD modelling uses an implementation given in ref. \cite{epperlein1986plasma}, in which $\mathbf{E}$ was numerically calculated from kinetic theory and then fitted with tabulated functions for the resistive ($\alpha_\parallel$, $\alpha_\perp$, $\alpha_\wedge$) and thermoelectric ($\beta_\parallel$, $\beta_\perp$, $\beta_\wedge$) transport coefficients. The transport coefficients describe how currents and heat flux are inhibited and deflected by the magnetic field \cite{froula2007quenching}. 

 These additional ExMHD effects are most important in high energy density (HED) plasmas such as Z-pinches \cite{haines2011review}, laser plasmas \cite{haines1986magnetic, froula2007quenching}, fast ignition fusion concepts \cite{nicolai2011effect}, dense fusion fuel hot-spots \cite{walsh2017self} and laser ablation fronts \cite{campbellmagnetic}. The ExMHD magnetic field advection can greatly exceed that due to the ideal advection with the fluid \cite{willingale2010fast}. Studies using the ExMHD codes Gorgon \cite{walsh2017self} and Hydra \cite{farmer2017simulation} found that heat insulation from self-generated magnetic fields can significantly change HED plasma temperature profiles. Accurate transport coefficients are therefore of considerable importance for plasma modelling.
 
 ExMHD results in an intricate set of feedback interactions. This includes, for example, growth of magnetic fields at the expense of fluid energy \cite{stamper1991review}, under processes such as the thermomagnetic instability \cite{tidman1974field, sherlock2020suppression}. The transport coefficients are also important for magnetic reconnection \cite{joglekar2014magnetic} in the weakly collisional plasma found in galaxy clusters and jets. Laboratory experiments emulating these magnetized jets \cite{liao2019design} and the turbulent dynamo process \cite{tzeferacos2018laboratory} also require ExMHD modelling.

In this work, we show that, rather than $\alpha_\perp$ and $\beta_\perp$, the primary quantities for magnetic transport are $\alpha_\perp-\alpha_\parallel$ and $\beta_\parallel-\beta_\perp$. Furthermore, the fits for $\alpha_\perp$ and $\beta_\perp$ given in ref. \cite{epperlein1986plasma} have the wrong dependence for weak magnetization, so they are not sufficiently accurate to calculate these quantities. This means that many ExMHD simulations in the literature, for example those using the Gorgon \cite{walsh2017self, campbellmagnetic}, CTC \cite{bissell2010field} and Hydra \cite{farmer2017simulation, davies2017laser} codes, have suffered inaccuracies and discontinuities in the magnetic transport. This then invalidates the thermal transport, indirectly affecting hydrodynamics. Using new, more accurate fit functions, we show that previous simulations have significantly over-estimated the cross-gradient Nernst advection and the resulting magnetic field dissipation. Recognizing the importance of $\alpha_\perp-\alpha_\parallel$ and $\beta_\parallel-\beta_\perp$, we re-cast the set of coefficients and thus reveal the inherent symmetry between the magnetic and heat transport, and the symmetry of the coefficients themselves.

The magnetic transport is described by the tensor ExMHD generalized Ohm's law, given by \cite{braginskii1958transport, epperlein1986plasma}
\begin{align}
    &\mathbf{E} = -\mathbf{u\times B} + \frac{\mathbf{J\times B}}{n_\mathrm{e}e} -\frac{\nabla .\underline P_\mathrm{e}}{n_\mathrm{e}e} + \frac{m_\mathrm{e}\underline \alpha.\mathbf{J}}{n_\mathrm{e}e^2\tau} - \frac{\underline\beta.\nabla T_\mathrm{e}}{e},\label{ohm}\\
    &\underline \alpha.\mathbf{J} =\alpha_\parallel(\mathbf{J}.\mathbf{\hat b})\mathbf{\hat b} + \mathbf{\hat b}\times(\alpha_\perp \mathbf{J\times \hat b} - \alpha_\wedge\mathbf{J}),\label{resistivity}\\
    &\underline \beta.\nabla T_\mathrm{e} =\beta_\parallel(\nabla T_\mathrm{e}.\mathbf{\hat b})\mathbf{\hat b} + \mathbf{\hat b}\times(\beta_\perp \nabla T_\mathrm{e}\mathbf{\times \hat b} + \beta_\wedge\nabla T_\mathrm{e}).\label{thermoelectric}
\end{align}
The first term in eqn. (\ref{ohm}) is the relativistic transformation from the ion fluid rest frame at velocity $\mathbf{u}$ and, taken alone, yields ideal MHD. The full Ohm's law also depends on the electron charge $-e$, mass $m_\mathrm{e}$, number density $n_\mathrm{e}$ and temperature $T_\mathrm{e}$. In quasi-neutral plasma $n_\mathrm{e}=\sum_j n_jZ_j$, where $n_j$ is the number density of ion species $j$ with ionization $Z_j$. There is also the Hall correction, written in terms of the current density $\mathbf{J}$. An electric field also arises due to gradients in the electron pressure tensor $\underline P_\mathrm{e}$. The inertial term has been neglected.

Coulomb collisions cause the appearance of the final two terms in eqn. (\ref{ohm}). The magnetic field causes resistivity to depend on the direction, such that the eqn. (\ref{resistivity}) must be decomposed into an orthogonal basis set parallel and perpendicular to the field direction $\mathbf{\hat b = B/|B|}$. Each component has its own dimensionless and positive transport coefficient $\alpha_\perp(\chi, \bar Z)$, $\alpha_\wedge(\chi, \bar Z)$ and $\alpha_\parallel(\bar Z)$ = $\alpha_\perp(0, \bar Z)$. Together these describe the magnetized deflection and inhibition of the plasma currents. Similarly, the collisional thermal force or thermoelectric term in eqn. (\ref{thermoelectric}) is driven by electron temperature gradients and depends on the coefficients $\beta_\perp(\chi, \bar Z)$, $\beta_\wedge(\chi, \bar Z)$ and $\beta_\parallel(\bar Z)$ = $\beta_\perp(0, \bar Z)$. These are functions of the average ion charge state $\bar Z=(\sum_j n_j Z_j^2)/(\sum_j n_j Z_j)$ and the dimensionless electron magnetization 
   \begin{align}
      \chi&= \frac{e|\mathbf{B}|\tau}{m_\mathrm{e}}=6\times 10^{16} \frac{|\mathbf{B}|T_\mathrm{eV}^{3/2}}{n_\mathrm{e}\bar Z\ln(\Lambda)},\label{chi}
\end{align}
where the electron Coulomb collision time is
\begin{align}
    \tau&=\frac{3\sqrt{\pi}}{4}\frac{4\pi\epsilon_0^2m_\mathrm{e}^2v_\mathrm{th}^3}{n_\mathrm{e}\bar Z e^4 \ln(\Lambda)}\label{tau}
   = 3.4\times 10^5 \frac{T_\mathrm{eV}^{3/2}}{n_\mathrm{e}\bar Z\ln(\Lambda)}\,\,\,\mathrm{s}.
   \end{align}
These expressions contain the electron-ion Coulomb logarithm (assumed to be $\ln(\Lambda)\gg 1$), the vacuum permittivity $\epsilon_0$ and the electron thermal speed $v_\mathrm{th}=\sqrt{2T_\mathrm{e}/m_\mathrm{e}}$. The numerical formulas are given in terms of $|\mathbf{B}|$ in Tesla, electron temperature $T_\mathrm{eV}$ in electron-volts and $n_\mathrm{e}$ in cm$^{-3}$. The magnetization $\chi$ gives the relative importance of gyro-motion and Coulomb collisions.

We now make the standard MHD assumption to retain only slow oscillations and therefore neglect displacement current, yielding $\mathbf{J}=c^2\epsilon_0\nabla\times\mathbf{B}$.  Following ref. \cite{walsh2020extended}, manipulation of eqns. (\ref{ohm}-\ref{thermoelectric}), using the vector components $\mathbf{J} = \mathbf{\hat b(J.\hat b)} + \mathbf{\hat b\times(J\times\hat b)}$, leads to the simplified form
\begin{align}
\mathbf{E} = &-\mathbf{u}_\mathrm B\mathbf{\times B} +D_\parallel\nabla\times\mathbf{B} - \frac{\nabla . \underline P_\mathrm{e}}{n_\mathrm{e}e}  - \frac{\beta_\parallel}{e}\nabla T_\mathrm{e},\label{ohm2}\\
\begin{split}
\mathbf{u}_\mathrm{B} = \,&\mathbf{u} -(1+\delta_\perp)\frac{\mathbf{J}}{n_\mathrm{e}e}+\delta_\wedge\frac{\mathbf{J\times\hat b}}{n_\mathrm{e}e} \\
&- \gamma_\perp\frac{\tau}{m_\mathrm{e}}\nabla T_\mathrm{e} + \gamma_\wedge\frac{\tau}{m_\mathrm{e}}\nabla T_\mathrm{e}\times\mathbf{\hat b},\label{ub}
\end{split}
\end{align}
where we have defined the magnetic advection velocity $\mathbf{u}_\mathrm B$ and the resistive magnetic diffusivity $D_\parallel=m_\mathrm{e}c^2\epsilon_0\alpha_\parallel/(n_\mathrm{e}e^2\tau)$. The required combinations of the $\alpha$ and $\beta$ coefficients motivate the definition of the new  transport coefficients \cite{walsh2020extended}
\begin{align}
&\delta_\perp(\chi, \bar Z) = \frac{\alpha_\wedge}{\chi},\qquad\qquad\gamma_\perp(\chi, \bar Z) = \frac{\beta_\wedge}{\chi},\label{perp}\\
&\delta_\wedge(\chi, \bar Z) = \frac{\alpha_\perp-\alpha_\parallel}{\chi},\quad\gamma_\wedge(\chi, \bar Z) = \frac{\beta_\parallel-\beta_\perp}{\chi}.\label{wedge}
\end{align}
The evolution of $\mathbf{B}$ is found via Faraday's law $\partial_t\mathbf{B}=-\nabla\times\mathbf{E}$. In order of appearance, the terms in eqn. (\ref{ohm2}) are then responsible for advection of $\mathbf{B}$ with velocity $\mathbf{u}_\mathrm B$, resistive diffusion of $\mathbf{B}$, the Biermann battery source term, and the Z-gradient source term \cite{haines1997saturation, sadler2020magnetization}. This form of Ohm's law has the advantage that the sole appearance of the ExMHD effects, that is, the $\perp$ and $\wedge$ coefficients, is within modifications to the magnetic advection velocity $\mathbf{u}_\mathrm{B}$ in eqn. (\ref{ub}). The coefficients $D_\parallel$ and $\beta_\parallel$ for the other terms in eqn. (\ref{ohm2}) are equivalent to those of the simpler resistive-MHD model and do not depend on $\mathbf{B}$.

In addition to the usual $D_\parallel$ resistive diffusion of magnetic field, the $\delta_\perp$ and $\delta_\wedge$ resistive terms alter the Hall velocity $-\mathbf{J}/(n_\mathrm{e}e)$ in eqn. (\ref{ub}), both in the parallel and transverse directions. The main effect of the thermal force is the Nernst advection \cite{nishiguchi1985nernst, haines1986heat} of $\mathbf{B}$  down the temperature gradient, with coefficient $\gamma_\perp$. There is also the $\gamma_\wedge$ cross-gradient Nernst advection \cite{walsh2020extended, davies2015importance} along isotherms. This cross-gradient term is important in HED plasmas \cite{walsh2019perturbation, walsh2020magnetized, farmer2017simulation}. 

We note that it is not the original $\alpha_\parallel$, $\alpha_\perp$, $\beta_\parallel$ and $\beta_\perp$ coefficients that are important for the magnetic transport in eqn. (\ref{ub}), but rather the differences between them. This is recognized in the definitions in eqn. (\ref{wedge}). However, we later show that the fit functions given in ref. \cite{epperlein1986plasma} are not sufficient to accurately calculate these differences.

The $\delta$ and $\gamma$ coefficients are fundamental in exposing the symmetry of the magnetic and thermal transport. This becomes apparent when eqn. (\ref{ub}) is compared with the equivalent expression from the electron heat flow \cite{epperlein1986plasma}
\begin{align}
&\mathbf{q}_\mathrm{e} =  - \frac{n_\mathrm{e}T_\mathrm{e}\tau}{m_\mathrm{e}}\underline \kappa.\nabla T_\mathrm{e} - \frac{T_\mathrm{e}}{e}\underline\beta.\mathbf{J},\label{heatflux}
\end{align}
The total electron energy flux, including the enthalpy flux and heat flow, is given by $U_\mathrm{e}\mathbf{u}_\mathrm{e} + \underline P_\mathrm{e}.\mathbf{u}_\mathrm{e} + \mathbf{q}_\mathrm{e}$, where $U_\mathrm{e}=m_\mathrm{e}n_\mathrm{e}|\mathbf{u}_\mathrm{e}|^2/2 + \mathrm{Tr}(\underline P_\mathrm{e})/2$ is the electron fluid energy density and $\mathbf{u}_\mathrm{e}=\mathbf{u}-\mathbf{J}/(n_\mathrm{e}e)$. Taking isotropic electron pressure with $\underline P_\mathrm{e}=n_\mathrm{e}T_\mathrm{e}\underline I$ and assuming $|\mathbf{u}_\mathrm{e}|\ll v_\mathrm{th}$, this total energy flux can be written as $n_\mathrm{e}T_\mathrm{e}\mathbf{ u_q}$, with
\begin{align}
    \begin{split}
\mathbf{u_q} = \,&\frac{5}{2}\mathbf{u} -\left(\frac{5}{2}+\beta_\perp\right)\frac{\mathbf{J}}{n_\mathrm{e}e} +\beta_\wedge\frac{\mathbf{J\times\hat b}}{n_\mathrm{e}e}\label{uq} \\&- \kappa_\perp\frac{\tau}{m_\mathrm{e}}\nabla T_\mathrm{e} + \kappa_\wedge\frac{\tau}{m_\mathrm{e}}\nabla T_\mathrm{e}\times\mathbf{\hat b}\\
&-(\beta_\parallel-\beta_\perp)\frac{(\mathbf{J.\hat b})}{n_\mathrm{e}e}\mathbf{\hat b} - (\kappa_\parallel-\kappa_\perp)\frac{\tau}{m_\mathrm{e}}(\mathbf{\hat b}.\nabla T_\mathrm{e})\mathbf{\hat b}.\raisetag{50pt}
\end{split}
\end{align}
Use of the $\delta$ and $\gamma$ transport coefficients now explicitly shows the symmetry between the magnetic flow [eqn. (\ref{ub})] and the electron energy flow [eqn. (\ref{uq})]. After replacing the $\delta$ and $\gamma$ coefficients with their $\beta$ and $\kappa$ counterparts, these expressions are almost equivalent. The only differences are the greater coefficient of $\mathbf{u}$ and the additional corrections along the field direction $\mathbf{\hat b}$ in eqn. (\ref{uq}), whereas magnetic advection along $\mathbf{B}$ is not possible.

It turns out that, by defining the $\delta$ and $\gamma$ coefficients to bring eqns. (\ref{ub}) and (\ref{uq}) into a symmetric form, the coefficients themselves also become symmetric. To show this, we must calculate them using eqns. (\ref{perp}-\ref{wedge}). The $\alpha_\perp$ and $\beta_\perp$ coefficients of Braginskii \cite{braginskii1958transport} result in $\lim_{\chi \to 0} \delta_\wedge=\lim_{\chi \to 0} \gamma_\wedge=0$. 
Epperlein and Haines (EH) \cite{epperlein1986plasma} later improved the coefficient dependencies for $\chi\rightarrow\infty$. However, equation (\ref{ub}) shows the importance of accurately calculating $\alpha_\perp-\alpha_\parallel$ and $\beta_\parallel-\beta_\perp$. This was not recognized in the EH fit functions, or in those of Ji and Held \cite{ji2013closure}. As a result, their approximations for $\alpha_\perp$ and $\beta_\perp$ imply that $\lim_{\chi \to 0} \delta_\wedge\neq 0$ and $\lim_{\chi \to 0} \gamma_\wedge\neq0$, in disagreement with Braginskii.

  \begin{figure}[t!]
  \includegraphics{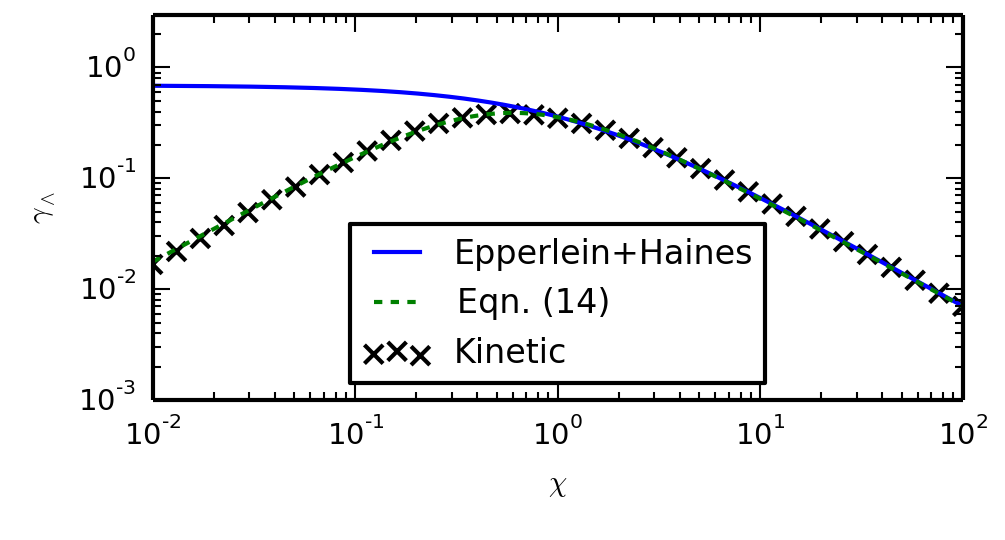}
  \caption{Plot of the $\gamma_\wedge$ cross-gradient Nernst transport coefficient for $\bar Z=1$, as calculated from the kinetic Fokker-Planck simulations. These results are accurately fitted with eqn. (\ref{gammacross}). Cross-gradient Nernst advection calculated from the fit functions of Epperlein and Haines \cite{epperlein1986plasma} is only accurate for $\chi>1$.  }
   \label{kinetic}
\end{figure}

  \begin{figure}[t!]
  \includegraphics{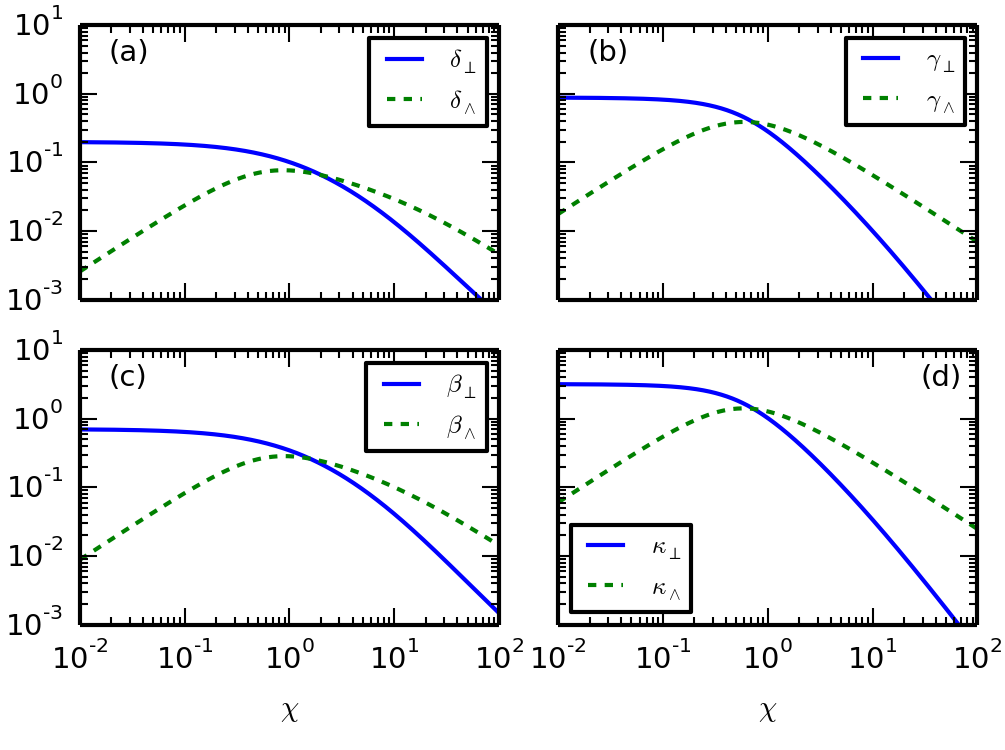}
  \caption{Plots of the symmetric transport coefficients for $\bar Z=1$. $\delta_\wedge$ and $\gamma_\wedge$ were calculated using eqns. (\ref{deltacross}-\ref{gammacross}), whereas all of the others can be accurately calculated using the results of ref.  \cite{epperlein1986plasma} and eqn. (\ref{perp}). (a) The Hall coefficients $\delta_\perp$ and $\delta_\wedge$. (b) The Nernst coefficients $\gamma_\perp$ and $\gamma_\wedge$. (c) The thermoelectric coefficients $\beta_\perp$ and $\beta_\wedge$. (d) The Spitzer coefficients $\kappa_\perp$ and $\kappa_\wedge$. }
   \label{coefsset}
\end{figure}

We now resolve this discrepancy and provide accurate fit functions. Our kinetic results follow the method of ref. \cite{epperlein1986plasma}, in which electrons are treated with the Fokker-Planck equation, with static ions. Furthermore, the electron distribution function is expanded \cite{thomas2012review, tzoufras2011vlasov} into its isotropic and anisotropic parts via $f_\mathrm{e}(\mathbf{v})=f_0(v) + \mathbf{v.f}_1(v)/v$, where $v=|\mathbf{v}|$. The truncation at first order is valid so long as $|\mathbf f_1|\ll|f_0|$. This limits the validity to plasma with shallow gradients, such that $v_\mathrm{th}\tau |\nabla T_\mathrm{e}|/T_\mathrm{e}\ll 1$ and $v_\mathrm{th}\tau |\nabla n_\mathrm{e}|/n_\mathrm{e}\ll 1$. This local assumption yields $f_0\simeq n_\mathrm{e}/(v_\mathrm{th}\sqrt{\pi})^3\exp(-v^2/v_\mathrm{th}^2)$, a Maxwellian at fixed density and temperature. Several authors \cite{luciani1983nonlocal, brodrick2018incorporating, henchen2018observation} have examined departures from this assumption. In a uniform plasma, $\mathbf{f}_1$ reaches a steady state given by
\begin{align}
    &\frac{e}{m_\mathrm{e}}\left(\mathbf{E}\frac{df_0}{dv} + \mathbf{B}\times\mathbf{f}_1\right)   -\frac{3\sqrt{\pi}}{4}\frac{v_\mathrm{th}^3}{v^3}\frac{\mathbf{f}_1}{\tau} + \mathbf C_\mathrm{ee}=0.\label{vfp}
\end{align}
The perturbation $\mathbf{f}_1$ reaches an equilibrium between the electromagnetic forces and the collision operators. The electron-ion collision operator in eqn. (\ref{vfp}) is a simple decay of $\mathbf{f}_1$ on a timescale $\tau$, whereas the electron-electron operator $\mathbf{C}_\mathrm{ee}$ is more complex and is given in ref. \cite{tzoufras2011vlasov}. 

Equation (\ref{vfp}) was solved numerically via an explicit iterative method, using fourth order numerical integrals and finite differences. The uniform velocity grid extended to $8v_\mathrm{th}$ with a resolution of $v_\mathrm{th}/15$. To isolate the $\alpha$ and $\beta$ coefficients, we assumed a fixed electric field and a transverse magnetic field. This yielded the steady state $\mathbf{f}_1$, which was numerically integrated \cite{thomas2012review} to find the resulting current $\mathbf{J} = -(4\pi e/3)\int_0^\infty \mathbf{f}_1v^3\,dv$ and heat flux $\mathbf{q}_\mathrm{e} = 5T_\mathrm{e}\mathbf{J}/(2e) +(2\pi m_\mathrm{e}/3)\int_0^\infty \mathbf{f}_1 v^5\,dv$. The $\alpha$ and $\beta$ coefficients are then found from eqns. (\ref{ohm}-\ref{resistivity}) and (\ref{heatflux}), using the fact that $\nabla T_\mathrm{e}=\mathbf{u}=0$. Finally, equations (\ref{perp}-\ref{wedge}) are used to calculate the $\delta$ and $\gamma$ coefficients. This process was repeated for different values of $\bar Z$ and $\mathbf{B}$. 
\begin{table*}[t]
\centering
\renewcommand{\arraystretch}{1.1}
\begin{tabular}{|l|llllllllllllll|}
\hline
$\bar Z$ & 1 & 2 & 3 & 4 & 5 & 6 & 7 & 8 & 10 & 12 & 14 & 20 & 30 & 60 \\ \hline
 $\alpha_\parallel$& 0.5061\, & 0.4295\, & 0.3950\, & 0.3750\, & 0.3618\, & 0.3524\, & 0.3454\, & 0.3399\, & 0.3319 \,& 0.3263\, & 0.3221 \,& 0.3144\, & 0.3081\, & 0.3015\, \\ %\hline
 $a_0$ & 3.8566 & 1.4509 & 0.8226 & 0.5975 & 0.4742 & 0.3997 & 0.3582 & 0.3214 & 0.2763 & 0.2450 & 0.2185 & 0.1857 & 0.1608 & 0.1374 \\ %\hline
 $a_1$& 4.8675 & 3.0454 & 2.8355 & 2.5790 & 2.4409 & 2.3423 & 2.2302 & 2.1812 & 2.0923 & 2.0465 & 2.0378 & 1.9555 & 1.8942 & 1.8310 \\ %\hline
 $a_2$& 9.7813 & 8.1847 & 7.4331 & 7.0947 & 6.8718 & 6.7199 & 6.6314 & 6.5429 & 6.4272 & 6.3389 & 6.2580 & 6.1518 & 6.0634 & 5.9722 \\ %\hline
 $\beta_\parallel$& 0.7029 & 0.9054 & 1.0180 & 1.0923 & 1.1456 & 1.1861 & 1.2180 & 1.2439 & 1.2834 & 1.3121 & 1.3341 & 1.3770 & 1.4139 & 1.4547 \\ %\hline
 $b_0$& 0.5589 & 0.1541 & 0.0792 & 0.0514 & 0.0381 & 0.0303 & 0.0254 & 0.0219 & 0.0176 & 0.0150 & 0.0133 & 0.0105 & 0.0086 & 0.0068 \\ %\hline
 $b_1$& 1.0599 & 0.5323 & 0.3880 & 0.3231 & 0.2831 & 0.2578 & 0.2398 & 0.2267 & 0.2082 & 0.1955 & 0.1868 & 0.1707 & 0.1582 & 0.1457 \\ %\hline
 $b_2$& 2.1643 & 1.6846 & 1.4931 & 1.3845 & 1.3173 & 1.2692 & 1.2336 & 1.2056 & 1.1655 & 1.1384 & 1.1180 & 1.0806 & 1.0505 & 1.0189 \\ \hline
\end{tabular}
\caption{Parameters for the fit functions of the $\delta_\wedge$ and $\gamma_\wedge$ transport coefficients presented in eqns. (\ref{deltacross}-\ref{gammacross}), as a function of ion charge $\bar Z$. The maximum error relative to the kinetic results is $10\%$ for $\delta_\wedge$ and $8\%$ for $\gamma_\wedge$.}
\label{parameters}
\end{table*}
\begin{figure}[t!]
  \includegraphics[width=2.55in]{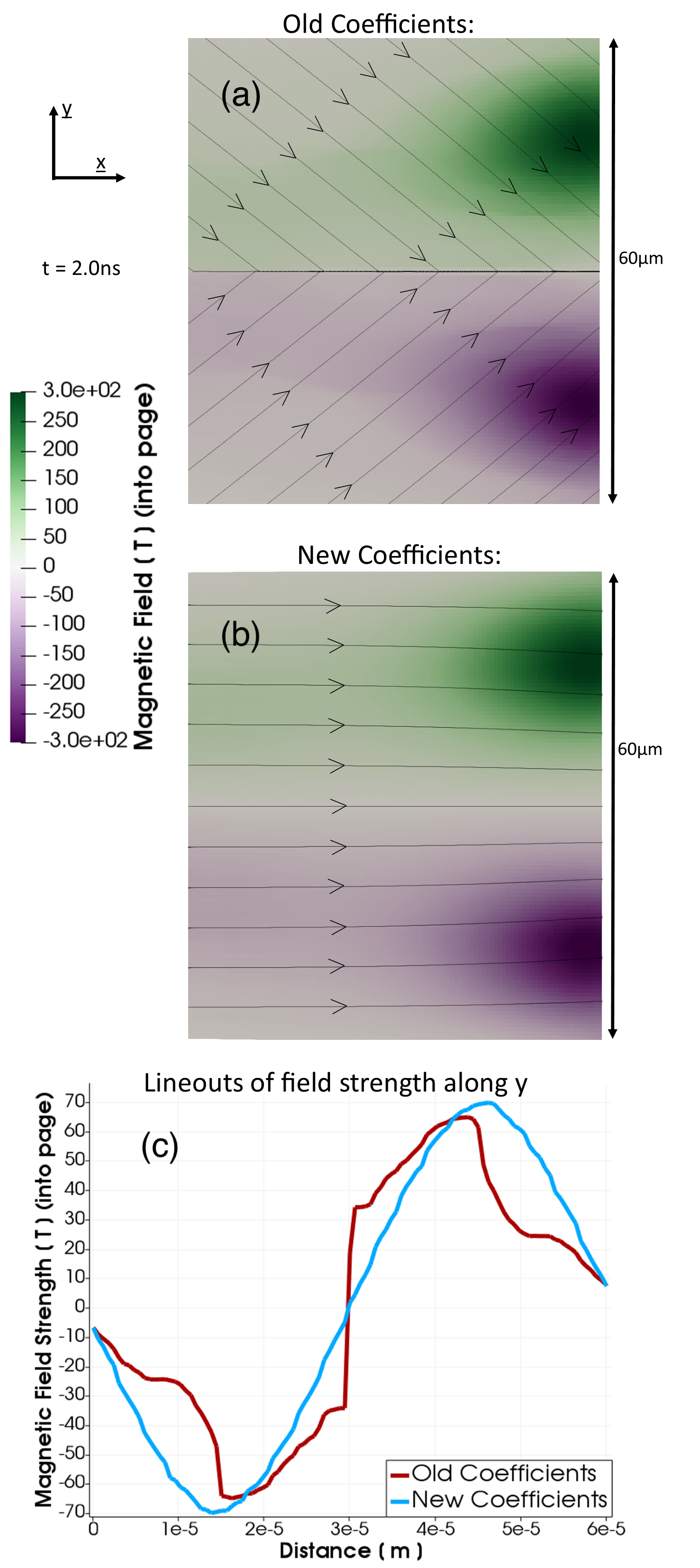}
  \caption{Magnetic field $B_z$ profiles from two equivalent two-dimensional ExMHD simulations after $2\,$ns. These used the coefficient fit functions of (a) Epperlein and Haines \cite{epperlein1986plasma} and (b) eqns. (\ref{deltacross}-\ref{gammacross}). Streamlines show the magnetic advection velocity from eqn. (\ref{ub}). (c) Line-outs of $B_z$, taken along $y$ at the center of the $x$ axis. }
   \label{simulation}
\end{figure}
The results for $\gamma_\wedge$ are presented in Fig. \ref{kinetic}. The kinetic results are plotted alongside the EH estimates \cite{epperlein1986plasma}. It is clear that their fit functions are only accurate for $\chi> 1$. The EH fits are sufficiently accurate to calculate $\delta_\perp$ and $\gamma_\perp$ with eqn. (\ref{perp}), but should not be used to calculate $\delta_\wedge$ and $\gamma_\wedge$ with eqn. (\ref{wedge}). More accurate fit functions for $\delta_\wedge$ and $\gamma_\wedge$, with the correct limits, are given by
\begin{align}
    \delta_\wedge(\chi, \bar Z) &= \frac{\chi + (1-\alpha_\parallel)\chi^2}{ a_0 +  a_1\chi+ a_2\chi^2+\chi^3},\label{deltacross}\\
    \gamma_\wedge(\chi, \bar Z) &= \frac{\chi + \beta_\parallel\chi^2}{ b_0 +  b_1\chi+ b_2\chi^2+\chi^3}.\label{gammacross}
\end{align}
Fig. \ref{kinetic} also shows the fit function (\ref{gammacross}). The $a_i$ and $b_i$ coefficients, presented in Table \ref{parameters}, were found via a least squares error minimization algorithm. The EH fit functions lead to inaccuracies in the $\delta_\wedge$ cross-Hall and $\gamma_\wedge$ cross-Nernst magnetic transport. Similarly, eqns. (\ref{deltacross}-\ref{gammacross}) are inaccurate if used to calculate $\alpha_\perp$ and $\beta_\perp$.

The full set of $\delta$, $\gamma$, $\beta$ and $\kappa$ symmetric transport coefficients are plotted in Fig. \ref{coefsset} for the case $\bar Z=1$. These coefficients, together with $\alpha_\parallel(\bar Z)$ and $\beta_\parallel(\bar Z)$, constitute a complete set. It is now obvious why we have labelled these the symmetric coefficients, since, in contrast to the (now defunct) $\alpha_\perp$ coefficient, all of them now have the same overall shape. By defining the $\delta$ and $\gamma$ coefficients to bring eqns. (\ref{ub}) and (\ref{uq}) into their symmetric form, the set of transport coefficients also becomes symmetric.

To assess time-dependent effects, the new fit functions were implemented in the ExMHD code Gorgon \cite{walsh2020extended}. A test problem was performed to recreate magnetic fields generated at the edge of an inertial confinement fusion hot-spot; understanding the magnetic dynamics is essential for assessing fuel thermal energy containment \cite{walsh2017self}. A density gradient between $50\,$gcm$^{-3}$ and $500\,$ gcm$^{-3}$ is set along the $x$ direction in a square $60 \,\mu$m Cartesian box, with $\bar Z=1$. The lower and upper $x$ boundaries were held at constant temperatures $2\,$keV and $1\,$keV, respectively. This results in a continual flux of heat from low to high $x$. Boundaries were periodic in $y$. A small sinusoidal temperature perturbation was initialized in $y$ such that the central plane is a fraction $1/60$ colder than the edges. Magnetic fields with $\chi<1$ are self-generated in the $z$ direction by the Biermann Battery mechanism and are predominantly advected by the Nernst and cross-gradient-Nernst velocities. The simulation is run for $2\,$ns with the results plotted in Fig. \ref{simulation}.

It is interesting to note that in the limit $\chi\rightarrow 0$, the EH fit functions \cite{epperlein1986plasma} predict $\lim_{\chi\rightarrow 0}\gamma_\wedge\neq 0$, giving a finite cross-gradient Nernst velocity $\simeq (\tau/m_\mathrm{e})\nabla T_\mathrm{e}\times\mathbf{\hat b}$. At spatial positions with $|\mathbf{B}|=0$, $\mathbf{\hat b}$ is undefined and so this predicts a discontinuity in $\mathbf{u}_\mathrm B$, shown by the convergence of streamlines in Fig. \ref{simulation}a. The new fit functions in eqns. (\ref{deltacross}-\ref{gammacross}), on the other hand, predict no such discontinuity in Fig. \ref{simulation}b. This artificial discontinuity also appears in the magnetic field profile line-outs shown in Fig. \ref{simulation}c. 

 The two magnetic field profiles differ significantly everywhere, not just at the discontinuity. Fig. \ref{simulation}a predicts a diagonal total Nernst advection in regions with $\chi\ll 1$, whereas the new fits in Fig. \ref{simulation}b predict a simple Nernst advection $\propto -\nabla T_\mathrm{e}$. There is only a slight cross-gradient Nernst velocity (in the $y$ direction) arising in regions with greater $|\mathbf{B}|$. This shows that, although the absolute least-squares errors of the $\alpha_\perp$ and $\beta_\perp$ EH fits are small, getting the correct functional form for $\chi<1$ is vitally important for the correct magnetic transport. The EH fit functions, resulting in this incorrect magnetic transport for $\chi<1$, have been widely implemented in several ExMHD codes \cite{davies2015importance, walsh2017self} since their inception.
 
 This miscalculation of cross-gradient Nernst advection has majorly impacted ExMHD simulations, and is therefore of more than just theoretical interest. Regions of positive and negative $B_z$ were artificially advected towards each other in Fig. \ref{simulation}a, causing reconnection and a reduction of the total flux. Fig. \ref{simulation}c shows that use of the accurate fits in eqns. (\ref{deltacross}-\ref{gammacross}) results in doubling of $|\mathbf{B}|$ and $\chi$ values in some regions. This means that in plasmas with dominant Nernst advection and $0.1<\chi<1$, ExMHD magnetic heat insulation was wrong by a factor of two or more in some regions.

Recent two-dimensional kinetic simulations of a laser ablation front \cite{hill2018enhancement} did not observe the diagonal Nernst advection behaviour predicted in Fig. \ref{simulation}a. In their simulations with $\chi< 0.1$, the cross-gradient Nernst velocity was three orders of magnitude less than the standard Nernst velocity, in agreement with Fig \ref{simulation}b. This remained true even in the denser plasma regions close to the target, where classical transport theory is expected to hold. A comparative lack of cross-gradient Nernst advection was also observed in kinetic simulations of the thermomagnetic instability, both with a Vlasov-Fokker-Planck \cite{sherlock2020suppression} and particle-in-cell \cite{PhysRevResearch.2.033233} approach.

In summary, we have shown that, once re-cast into a new set, all of the transport coefficients have the same behavior. This elucidates the symmetry of the magnetic and thermal transport in a collisional magnetized plasma. To accurately calculate magnetic transport for $\chi<1$, the fit functions of Epperlein and Haines \cite{epperlein1986plasma} must be updated. These previous fit functions massively over-estimated the cross-Nernst and cross-Hall advection, causing artificial magnetic discontinuities and dissipation. The new fits also explain the apparent discrepancies between kinetic simulations \cite{hill2018enhancement} and ExMHD simulations in the literature. This more natural and accurate description of magnetic transport will improve modelling capabilities for a wide range of magnetized HED plasma experiments. 

\begin{acknowledgments}
Research presented in this article was supported by Los Alamos National Laboratory (LANL) under Laboratory Directed Research and Development project number 20180040DR and the Center for Nonlinear Studies. The work was also performed under the auspices of the U.S. Department of Energy by Lawrence Livermore National Laboratory under Contract DE-AC52-07NA27344. This document was prepared as an account of work sponsored by an agency of the United States government. Neither the United States government nor Lawrence Livermore National Security, LLC, nor LANL, nor any of their employees makes any warranty, expressed or implied, or assumes any legal liability or responsibility for the accuracy, completeness, or usefulness of any information, apparatus, product, or process disclosed, or represents that its use would not infringe privately owned rights. Reference herein to any specific commercial product, process, or service by trade name, trademark, manufacturer, or otherwise does not necessarily constitute or imply its endorsement, recommendation, or favoring by the United States government, LANL, or Lawrence Livermore National Security, LLC. The views and opinions of authors expressed herein do not necessarily state or reflect those of the United States government, LANL or Lawrence Livermore National Security, LLC, and shall not be used for advertising or product endorsement purposes.
\end{acknowledgments}
%\bibliography{references.bib}

\begin{thebibliography}{10}

\bibitem{braginskii1958transport}
S.~I. Braginskii,
\newblock Sov. Phys. JETP {\bfseries 6}, 358 (1958).

\bibitem{epperlein1986plasma}
E.~M. Epperlein and M.~G. Haines,
\newblock Phys. Fluids {\bfseries 29}, 1029 (1986).

\bibitem{froula2007quenching}
D.~H. Froula, J.~S. Ross, B.~B. Pollock, P.~Davis, A.~N. James, L.~Divol, M.~J.
  Edwards, A.~A. Offenberger, D.~Price, R.~P.~J. Town, {\em et~al.},
\newblock Phys. Rev. Lett. {\bfseries 98}, 135001 (2007).

\bibitem{haines2011review}
M.~G. Haines,
\newblock Plasma Phys. Cont. Fusion {\bfseries 53}, 093001 (2011).

\bibitem{haines1986magnetic}
M.~G. Haines,
\newblock Canadian J. Phys. {\bfseries 64}, 912 (1986).

\bibitem{nicolai2011effect}
P.~Nicola{\"\i}, J.-L. Feugeas, C.~Regan, M.~Olazabal-Loum{\'e}, J.~Breil,
  B.~Dubroca, J.-P. Morreeuw, and V.~Tikhonchuk,
\newblock Phys. Rev. E {\bfseries 84}, 016402 (2011).

\bibitem{walsh2017self}
C.~A. Walsh, J.~P. Chittenden, K.~McGlinchey, N.~P.~L. Niasse, and B.~D.
  Appelbe,
\newblock Phys. Rev. Lett. {\bfseries 118}, 155001 (2017).

\bibitem{campbellmagnetic}
P.~T. Campbell, C.~A. Walsh, J.~P. Chittenden, A.~Crilly, G.~Fiksel, P.~M.
  Nilson, B.~K. Russell, A.~G.~R. Thomas, K.~Krushelnick, and L.~Willingale, 
  \newblock Phys. Rev. Lett.{},  (Accepted 2020).

\bibitem{willingale2010fast}
L.~Willingale, A.~G.~R. Thomas, P.~M. Nilson, M.~C. Kaluza, S.~Bandyopadhyay,
  A.~E. Dangor, R.~G. Evans, P.~Fernandes, M.~G. Haines, C.~Kamperidis, {\em
  et~al.},
\newblock Phys. Rev. Lett. {\bfseries 105}, 095001 (2010).

\bibitem{farmer2017simulation}
W.~A. Farmer, J.~M. Koning, D.~J. Strozzi, D.~E. Hinkel, L.~F. Berzak~Hopkins,
  O.~S. Jones, and M.~D. Rosen,
\newblock Phys. Plasmas {\bfseries 24}, 052703 (2017).

\bibitem{stamper1991review}
J.~A. Stamper,
\newblock Laser and Particle Beams {\bfseries 9}, 841 (1991).

\bibitem{tidman1974field}
D.~A. Tidman and R.~A. Shanny,
\newblock Phys. Fluids {\bfseries 17}, 1207 (1974).

\bibitem{sherlock2020suppression}
M.~Sherlock and J.~J. Bissell,
\newblock Phys. Rev. Lett. {\bfseries 124}, 055001 (2020).

\bibitem{joglekar2014magnetic}
A.~S. Joglekar, A.~G.~R. Thomas, W.~Fox, and A.~Bhattacharjee,
\newblock Phys. Rev. Lett. {\bfseries 112}, 105004 (2014).

\bibitem{liao2019design}
A.~S. Liao, S.~Li, H.~Li, K.~Flippo, D.~Barnak, K.~V. Kelso,
  C.~Fiedler~Kawaguchi, A.~Rasmus, S.~Klein, J.~Levesque, {\em et~al.},
\newblock Phys. Plasmas {\bfseries 26}, 032306 (2019).

\bibitem{tzeferacos2018laboratory}
P.~Tzeferacos, A.~Rigby, A.~F.~A. Bott, A.~R. Bell, R.~Bingham, A.~Casner,
  F.~Cattaneo, E.~M. Churazov, J.~Emig, F.~Fiuza, {\em et~al.},
\newblock Nat. Comms. {\bfseries 9}, 591 (2018).

\bibitem{bissell2010field}
J.~J. Bissell, C.~P. Ridgers, and R.~J. Kingham,
\newblock Phys. Rev. Lett. {\bfseries 105}, 175001 (2010).

\bibitem{davies2017laser}
J.~R. Davies, D.~H. Barnak, R.~Betti, E.~M. Campbell, P.-Y. Chang, A.~B.
  Sefkow, K.~J. Peterson, D.~B. Sinars, and M.~R. Weis,
\newblock Phys. Plasmas {\bfseries 24}, 062701 (2017).

\bibitem{walsh2020extended}
C.~A. Walsh, J.~P. Chittenden, D.~W. Hill, and C.~Ridgers,
\newblock Phys. Plasmas {\bfseries 27}, 022103 (2020).

\bibitem{haines1997saturation}
M.~G. Haines,
\newblock Phys. Rev. Lett. {\bfseries 78}, 254 (1997).

\bibitem{sadler2020magnetization}
J.~D. Sadler, H.~Li, and B.~M. Haines,
\newblock Phys. Plasmas {\bfseries 27}, 072707 (2020).

\bibitem{nishiguchi1985nernst}
A.~Nishiguchi, T.~Yabe, and M.~G. Haines,
\newblock Phys. Fluids {\bfseries 28}, 3683 (1985).

\bibitem{haines1986heat}
M.~G. Haines,
\newblock Plasma Phys. Cont. Fusion {\bfseries 28}, 1705 (1986).

\bibitem{davies2015importance}
J.~R. Davies, R.~Betti, P.-Y. Chang, and G.~Fiksel,
\newblock Phys. Plasmas {\bfseries 22}, 112703 (2015).

\bibitem{walsh2019perturbation}
C.~A. Walsh, K.~McGlinchey, J.~K. Tong, B.~D. Appelbe, A.~Crilly, M.~F. Zhang,
  and J.~P. Chittenden,
\newblock Phys. Plasmas {\bfseries 26}, 022701 (2019).

\bibitem{walsh2020magnetized}
C.~A. Walsh, A.~J. Crilly, and J.~P. Chittenden,
\newblock Nuclear Fusion {\bfseries 60}, 106006 (2020).

\bibitem{ji2013closure}
J.-Y. Ji and E.~D. Held,
\newblock Phys. Plasmas {\bfseries 20}, 042114 (2013).

\bibitem{thomas2012review}
A.~G.~R. Thomas, M.~Tzoufras, A.~P.~L. Robinson, R.~J. Kingham, C.~P. Ridgers,
  M.~Sherlock, and A.~R. Bell,
\newblock J. Comp. Phys. {\bfseries 231}, 1051 (2012).

\bibitem{tzoufras2011vlasov}
M.~Tzoufras, A.~R. Bell, P.~A. Norreys, and F.~S. Tsung,
\newblock J. Comp. Phys. {\bfseries 230}, 6475 (2011).

\bibitem{luciani1983nonlocal}
J.~F. Luciani, P.~Mora, and J.~Virmont,
\newblock Phys. Rev. Lett. {\bfseries 51}, 1664 (1983).

\bibitem{brodrick2018incorporating}
J.~P. Brodrick, M.~Sherlock, W.~A. Farmer, A.~S. Joglekar, R.~Barrois,
  J.~Wengraf, J.~J. Bissell, R.~J. Kingham, D.~Del~Sorbo, M.~P. Read, {\em
  et~al.},
\newblock Plasma Phys. Cont. Fusion {\bfseries 60}, 084009 (2018).

\bibitem{henchen2018observation}
R.~J. Henchen, M.~Sherlock, W.~Rozmus, J.~Katz, D.~Cao, J.~P. Palastro, and
  D.~H. Froula,
\newblock Phys. Rev. Lett. {\bfseries 121}, 125001 (2018).

\bibitem{hill2018enhancement}
D.~W. Hill and R.~J. Kingham,
\newblock Phys. Rev. E {\bfseries 98}, 021201(R) (2018).

\bibitem{PhysRevResearch.2.033233}
K.~M. Schoeffler and L.~O. Silva,
\newblock Phys. Rev. Research {\bfseries 2}, 033233 (2020).

\end{thebibliography}
%\bibliographystyle{h-physrev}

\end{document}